\documentstyle [twocolumn,prb,aps,floats,epsfig] {revtex}

\begin{document}
\draft
\flushbottom
\begin{title}
{\bf Decay channels and appearance sizes of doubly anionic gold and 
silver clusters}
\end{title} 
\author{Constantine Yannouleas and Uzi Landman} 
\address{
School of Physics, Georgia Institute of Technology,
Atlanta, Georgia 30332-0430 }
\date{January 2000}
\maketitle
\begin{abstract}
Second electron affinities of Au$_N$ and Ag$_N$ clusters and the dissociation 
energies for fission of the Au$_N^{2-}$ and Ag$_N^{2-}$ dianions are
calculated using the finite-temperature shell-correction method and allowing 
for triaxial deformations. Dianionic clusters with $N > 2$ are found to be
energetically stable against fission, leaving electron autodetachment as the
dominant decay process. The second electron affinities exhibit pronounced shell
effects in excellent agreement with measured abundance spectra for Au$_N^{2-}$
($N < 30$), with appearance sizes $n_a^{2-}$(Au)$=12$ and $n_a^{2-}$(Ag)$=24$.
\end{abstract}
\pacs{Pacs Numbers: 36.40.Wa, 36.40.Qv, 36.40.Cg}
\narrowtext

Unlike the case of multiply charged cationic species,
the production and observation of gas-phase doubly anionic aggregates
had remained for many years a challenging experimental goal. However,
with the availability of large carbon clusters (which can easily
accomodate the repulsion between the two excess electrons) this state of 
affairs changed, including observation of doubly negative fullerenes,
\cite{comp1} $C_{60}^{2-}$, and fullerene derivatives, \cite{comp2} 
as well as a recent measurement of the photoelectron spectrum
of the citric acid dianion. \cite{wang} Moreover, such observations are not 
limited to carbon based aggregates and organic molecules, with
a first observation of doubly anionic metal clusters (specifically gold 
clusters) reported \cite{schw1,schw2} most recently. A few theoretical studies
of multiply charged anionic fullerenes 
\cite{yann1,pede} and alkali-metal (sodium) clusters \cite{yann2} have also 
appeared, but overall the field of multiply anionic aggregates remains at 
an embryonic stage.

In this paper, we investigate the stability and decay channels of Au$_N^{2-}$ 
and Ag$_N^{2-}$ at finite temperature, and determine their appearance sizes
$n^{2-}_a$ (clusters with $N < n^{2-}_a$ are energetically unstable).  
Two decay channels of doubly anionic clusters need to be considered: 
(i) binary fission,
\begin{equation}
M^{2-}_N \rightarrow M^-_P + M^-_{N-P}~,
\label{eq1}
\end{equation}
which has a well known analog in the case of doubly cationic clusters,
\cite{brec,bjor,yann3} and (ii) electron autodetachment via emission through a 
Coulombic barrier, \cite{yann2} 
\begin{equation}
M^{2-}_N \rightarrow M^-_N + e~,
\label{eq2}
\end{equation}
with an analogy to proton and alpha decay in atomic nuclei. \cite{naza,book}
The theoretical approach we use is a finite-temperature semi-empirical
shell-correction method (SCM), which incorporates triaxial
shapes and which has been previously used successfully to describe 
the properties of neutral and cationic metal clusters. \cite{yann4}
Our main conclusion is that, unlike the case of doubly cationic metal clusters,
\cite{brec,yann3} fission of Au$^{2-}_N$ and Ag$^{2-}_N$ is not a dominant 
process, and that the appearance sizes of these doubly anionic clusters
are determined by electron autodetachment. Our results for the second electron
affinities exhibit pronounced electronic shell effects and are in excellent
agreement with most recent experimental data \cite{schw2} for Au$^{2-}_N$ with
$n^{2-}_a=12$. For Ag$^{2-}_N$, we predict $n^{2-}_a=24$.

The finite-temperature multiple electron affinities of a cluster of $N$
atoms of valence $v$ (we take $v=1$ for Au and Ag) are defined as
\begin{equation}
A_Z(N,\beta)=F(\beta, vN, vN+Z-1)-F(\beta,vN,vN+Z)~,
\label{eq3}
\end{equation}
where $F$ is the free energy, $\beta=1/k_B T$, and $Z \geq 1$ is
the number of excess electrons in the cluster (e.g., the first and second
affinities correspond to $Z=1$ and $Z=2$, respectively). To determine the free
energy, we use the shell correction method. In the SCM, $F$ is separated into 
a smooth liquid-drop-model (LDM) part $\widetilde{F}_{\text{LDM}}$ (varying 
monotonically with $N$), and a Strutinsky-type shell-correction term 
$\Delta F_{\text{sp}}=
F_{\text{sp}}-\widetilde{F}_{\text{sp}}$, where $F_{\text{sp}}$ is the 
canonical (fixed $N$ at a given $T$) free energy of the valence electrons, 
treated as independent single particles moving in an effective mean-field 
potential (approximated by a modified Nilsson hamiltonian pertaining to 
triaxial cluster shapes),
and $\widetilde{F}_{\text{sp}}$ is the Strutinsky-averaged free energy.
The smooth $\widetilde{F}_{\text{LDM}}$ contains volume, surface, and
curvature contributions, whose coefficients are determined as described
in Ref.\ \onlinecite{yann4}, with experimental values and temperature 
dependencies. In addition to the finite-temperature contribution due to the 
electronic entropy, the entropic contribution from thermal shape fluctuations 
is evaluated via a Boltzmann averaging. \cite{yann4}$^{(a)}$

We note here that the smooth contribution $\widetilde{A}_Z(N,\beta)$ to the 
full multiple electron affinities ${A}_Z(N,\beta)$ can be approximated 
\cite{yann2} by the LDM expression
\begin{equation}
\widetilde{A}_Z = \widetilde{A}_1 -\frac{(Z-1)e^2}{R(N)+\delta_0}
= W -\frac{(Z-1+\gamma)e^2}{R(N)+\delta_0}~,
\label{eq4}
\end{equation}
where $R(N)=r_s N^{1/3}$ is the radius of the positive background ($r_s$ is
the Wigner-Seitz radius which depends weakly on $T$ due to volume dilation),
$\gamma=5/8$, $\delta_0$ is an electron spillout parameter, and the work 
function $W$ is assumed to be temperature independent [we take $W$(Au)$=5.31$ 
eV and $W$(Ag)$=4.26$ eV]. 

In a recent experiment, \cite{schw2} singly anionic gold clusters
Au$_N^-$ ($N \leq 28$) were stored in a Penning trap, size selected, and
transformed into dianions, Au$_N^{2-}$, through irradiation by an electron 
beam. The measured \cite{schw2} relative intensity ratios of the dianions to 
\begin{figure}[t]
\centering\epsfig{file=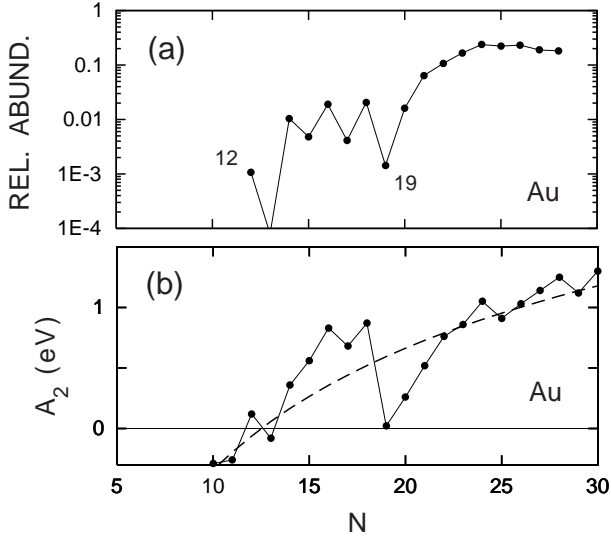,%
width=8.1cm,clip=,angle=0}
\caption{
(a) Measured [see figure 3(a) in Ref.\ \protect\onlinecite{schw2}] average 
relative abundances of Au$_N^{2-}$ clusters (i.e., the ratio of the
number of the observed dianions over the sum of the numbers of corresponding 
singly-anionic precursors and dianions) as a function of cluster size. Note 
the logarithmic ordinate scaling. (b) Calculated second electron affinity
($A_2$ in eV) for gold clusters at $T=300$ K plotted versus $N$. Results
from SCM calculations are connected by a solid line, and LDM
results [see Eq.\ (\ref{eq4}) with $Z=2$] are depicted by the dashed line.
$A_2 > 0$ corresponds to stable dianionic Au$_N^{2-}$ clusters; note the
appearance size $n_a^{2-}=12$. Energies in units of eV.
}
\end{figure}
their monoanionic  precursors are reproduced in Fig.\ 1(a); they exhibit 
size-evolutionary patterns (arising from electronic shell effects) 
reminiscent of those found earlier in the mass abundance spectra, ionization 
potentials and first electron affinities of alkali- and coinage-metal clusters.
\cite{note1} Since the stability of the dianions relative to 
their monoanionic precursors depends on the second electron affinity $A_2$, it
may be expected that $A_2$ and the relative signal intensity of the 
Au$_N^{2-}$ clusters will exhibit correlated patterns as a function of size. 
Here we note that stable dianions must have $A_2 > 0$, whereas those with 
$A_2 < 0$ are unstable and decay via process (ii), i.e., via electron emission
through a Coulombic barrier \cite{yann2} (see below).

In Fig.\ 1(b), we display the SCM theoretical results \cite{note2,note3} for 
the second electron affinity of Au$_N$ clusters in the size range 
$ 10 \leq N \leq 30$. These results correlate remarkably well with the measured
relative abundance spectrum [see Fig.\ 1(a)]. Note in particular: (i) the 
observed and predicted appearance size n$_a^{2-}$(Au)$=12$; (ii) the relative
instability of Au$_{13}^{2-}$ [portrayed by its absence in Fig.\ 1(a) and the
negative $A_2$ value in Fig.\ 1(b) associated with the closing of a
spheroidal electronic subshell (containing 14 electrons) in the singly anionic
Au$_{13}^-$ parent cluster, see Ref.\ \onlinecite{yann4}(b)]; (iii) the 
pronounced lower stability of Au$_{19}^{2-}$ relative to its neighboring 
cluster sizes [associated with the closing of a major electronic shell 
(containing 20 electrons) in the Au$_{19}^-$ parent cluster]; (iv) 
the overall similarity between
the trends in Fig.\ 1(a) and Fig.\ 1(b)(that is, even-odd alternations for
$N \leq 19$ with a sole discrepancy at $N=15$, and the monotonic behavior for
$N \geq 19$). Underlying the pattern shown in Fig.\ 1(b) are electronic
shell effects [compare in Fig. 1(b) the shell-corrected results indicated by
the solid line with the LDM curve] combined with energy-lowering shape 
deformations of the clusters (which are akin to Jahn-Teller distortions and 
are associated with the lifting of spectral degeneracies for open-shell 
cluster sizes).

\begin{figure}[t]
\centering\epsfig{file=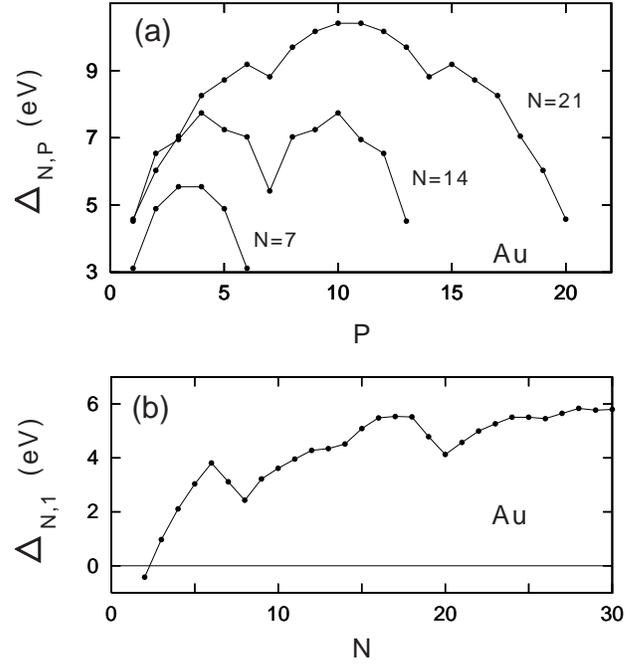,%
width=8.1cm,clip=,angle=0}
\caption{
(a) Fission dissociation energies ($\Delta_{N,P}$ in eV) for binary fission
Au$_N^{2-} \rightarrow $Au$_P^- + $Au$_{N-P}^-$, calculated at $T=300$ K with 
the SCM for parent dianionic clusters with $N=7$, 14 and 21, and plotted 
versus $P$. Note that in all cases the most favorable fission channel 
corresponds to $P=1$. (b) SCM fission dissociation energies, $\Delta_{N,1}$, 
at $T=300$ K for the most favorable channel, plotted versus cluster size. 
Exothermic fission ($\Delta_{N,1} < 0$) is found only for the smallest cluster.
}
\end{figure}
To explore the energetic stability of the Au$_N^{2-}$ clusters against
binary fission [see Eq.\ (\ref{eq1})], we show in Fig.\ 2(a) SCM results, at 
selected cluster sizes ($N=7, 14$ and 21), for the fission dissociation 
energies $\Delta_{N,P}= F$(Au$^-_P)+F($Au$^-_{N-P})-F($Au$^{2-}_N)$, with the 
total free energies of the parent dianion and the singly-charged fission 
products calculated at $T=300$ K.
For all Au$^{2-}_N$ parent clusters, the energetically
favorable channel (lowest $\Delta_{N,P}$) corresponds to $P=1$ (i.e., one of
the fission products is the closed-shell Au$^-$ anion). The influence of
shell effects on the fission dissociation energies is evident particularly
in cases where the fission channel involves closed-shell magic products
(see $P=7$, and equivalently $P=14$, for $N=21$, and 
the pronounced effect at $P=7$ for $N=14$ where
both fission products are magic). The fission results summarized in Fig.\ 2(b)
for the most favorable channel ($P=1$) illustrate that exothermic fission
(that is $\Delta_{N,P} < 0$) is predicted to occur only for the smallest
size ($N=2$). This, together with the existence of a fission barrier, leads us
to conclude that the decay of Au$_N^{2-}$ clusters is dominated by the 
electron autodetachment process (which is operative when $A_2 < 0$ and 
involves tunneling through a Coulomb barrier \cite{yann2}), rather than by
fission.

\begin{figure}[t]
\centering\epsfig{file=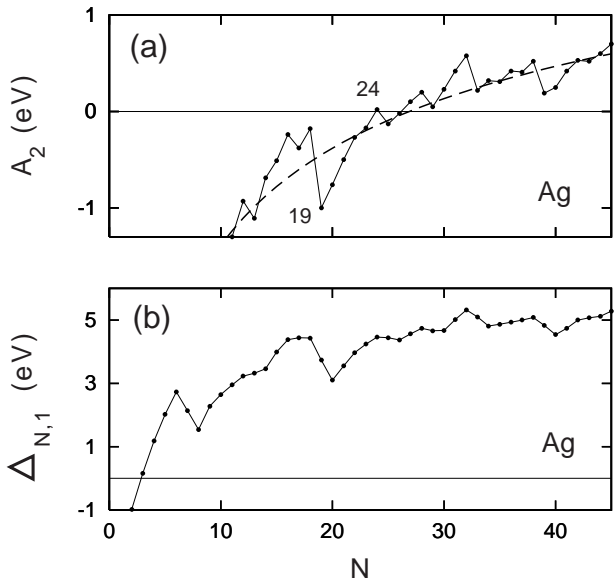,%
width=8.1cm,clip=,angle=0}
\caption{
SCM second electron affinities [$A_2$ in (a)] and fission dissociation energies
[$\Delta_{N,1}$ in (b)] for the most favorable channel ($P=1$) for Ag$_N^{2-}$
clusters at $T=300$ K, plotted versus cluster size. In (a) LDM results [see
Eq.\ (\ref{eq4}) with $Z=2$] are depicted by the dashed line. Note the 
appearance size $n_a^{2-}=24$. Energies in units of eV.
}
\end{figure}
Finally, we show in Fig.\ 3 SCM results for the second electron affinity 
[$A_2$ in Fig.\ 3(a)] and the fission dissociation energies [$\Delta_{N,P}$ in
Fig.\ 3(b)] corresponding to the most favorable channel $(P=1)$ for 
silver dianionic clusters Ag$_N^{2-}$. Again binary fission is seen to be
endothermic except for $N=2$, and the appearance size for Ag$_N^{2-}$
(i.e., the smallest size with $A_2 > 0$) is predicted to be 
$n_a^{2-}$(Ag)$=24$. The shift of the appearance size to a larger value
than that found for gold dianionic clusters [that is $n_a^{2-}$(Au)$=12$,
see above] can be traced to the smaller
work function of silver, as can be seen from the LDM curves calculated 
through the use of Eq.\ (\ref{eq4}) with $Z=2$. \cite{note4} ~~\\
~~~~~\\
This research was supported by a grant from the U.S. Department of Energy
(Grant No. FG05-86ER45234).


\begin{references}
\bibitem{note2}
For the rather small Au$_N$ and Ag$_N$ clusters discussed here ($N \leq 30$
and $N \leq 40$, respectively), 
our results at $T=0$ K and $T=300$ K differ only slightly. Since, however,
the experiments are carried out at finite temperatures, we present here
the $T=300$ K results for the sake of completeness. For cases where the SCM
reveals significant thermal effects portrayed by reduction and smearing out of
electronic shell effects (e.g., for higher temperatures and/or larger $r_s$'s 
as in the case of Na$_N$ and K$_N$ clusters), see Ref.\ \onlinecite{yann4}(a).
\bibitem{note4}
The use of a finite value [see Ref.\ \onlinecite{yann4}(b)] for the spillout 
parameter $\delta_0$ in the case of negatively charged clusters is crucial
for obtaining accurate results. Taking $\delta_0=0$ (as was done in Ref.\ 
\onlinecite{schw2}) yields a higher value for the appearance 
sizes.
\end{references}
\end{document}